\documentclass{ws-p8-50x6-00}

\begin{document}

\title{Recent Results and Current Status of BES}

\author{XU Guofa\\
    (Representing BES Collaboration)}

\address{Institute of High Energy Physics(IHEP)\\
         Chinese Academy of Sciences(CAS)\\
         Beijing 100039, P. R. China\\
         E-mail: xugf@mail.ihep.ac.cn}


\maketitle
\abstracts{The current status of BESII is reported.  
Some published results based on the data collected by BESI and some new 
preliminary results based on the BESII data are reviewed in this paper.}
\section{Introduction}
\begin{figure}[h]
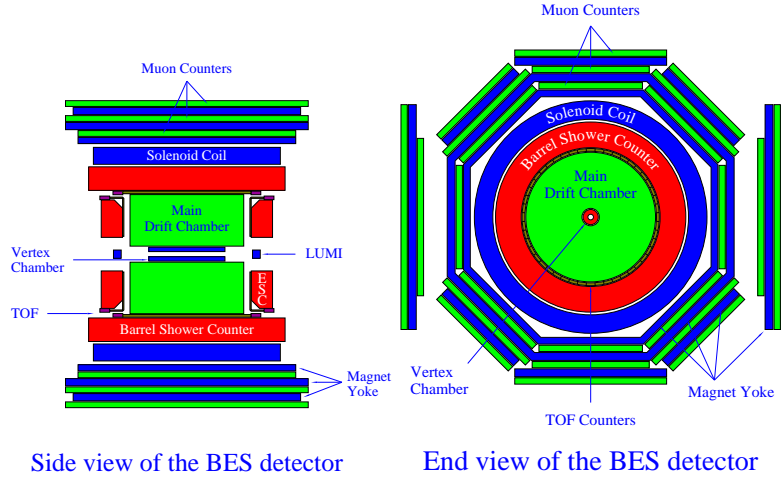

\begin{center}
\epsfxsize=12pc 
\epsfbox{BES2_view2.epsi}
\epsfxsize=12pc 
\epsfbox{BES2_view1.epsi}  
\caption{The BESII Detector  \label{fig1}}
\end{center}
\end{figure}

The Beijing Electron Positron Collider(BEPC) covers the  
center-of-mass energy range from 2.0 to 5.0 GeV .
Beijing Spectrometer(BES) is a large general purpose solenoidal 
detector at BEPC and is described in detail in Ref. 1. Both  
BEPC and BES were upgraded from 1995 to 1997, 
the upgrade of BEPC included 
moving the insertion quadrupoles closer to the interaction point and 
increasing the total RF voltage. 
There was a factor of $1.5 \sim 2.0$ improvement in the luminosity,  now 
the luminosity at the $J/\psi$ energy is around 
$(4 - 5)\times 10^{30}$ cm$^{-2}$s$^{-1}$. The hadronic event rate 
recorded at $J/\psi$ peak can now reach $6\sim 8$ Hz. 
The performance parameters of the upgraded BES(BESII) are  listed in 
Table~\ref{tab1}, and the structure of BESII is shown 
in Fig.~\ref{fig1}.
\par
\begin{table}[bh]
\caption{ The Performance Parameters of BES \label{tab1}}
\begin{center}
\begin{tabular}{|c|c|c|}
\hline
Detector  & Major parameter            & BESII \\\hline
VC        & $\sigma_{xy}$($\mu m$)      & 100  \\\hline
          & $\sigma_{xy}$($\mu m$)      & 190-220 \\
MDC       & $\Delta p/p$ ($\%$)        & $1.78\sqrt{1+p^2}$ \\
          & $\sigma_{dE/dx}$ ($\%$)    & 8.4                \\\hline
BTOF      & $\sigma_T$ (ps)            &  180   \\
          & $L_{atten}$ (m)            & 3.5 - 5.5  \\\hline
ETOF      & $\sigma_T$ (ps)            & 720   \\\hline
BSC       & $\Delta E/\sqrt{E}$ ($\%$) & $23\%$ \\
          & $\sigma_{z}$(cm)           & 2.3 \\\hline
ESC       & $\Delta E/\sqrt{E}$ ($\%$) & $21\%$ \\\hline
$\mu$ counter & $\sigma_{z}$(cm)       & 5.5 \\\hline
DAQ       & dead time (ms)             &  8  \\\hline
\end{tabular}
\end{center}
\end{table}
\par
\par

\section{Recent Results}
\subsection{R measurement}
   R, one of the most fundamental quantities in particle physics, is 
defined as 
\begin{equation}
   R\equiv \frac{\sigma(e^+e^-\to hadrons)}{\sigma(e^+e^-\to \mu^+ 
          \mu^-)}
\label{eq1}
\end{equation}
   On the other hand, the R values experimentally  can be estimated using  
 the following equation(Eq.(~\ref{eq1}))
\begin{equation}
   R= \frac{1}{\sigma(e^+e^-\to \mu^+ \mu^-)}\cdot 
      \frac{N_{had}-N_{bg}}{\pounds \cdot \varepsilon_{had}\cdot 
          (1+\delta)},
\label{eq1}
\end{equation}
\noindent
 where $N_{had}$ is the number of observed hadronic events, 
 $N_{bg}$ is the number of background events, $\pounds$ is the integrated
luminosity, $\varepsilon_{had}$ is the detection efficiency for $N_{had}$, 
and $\delta$ is the radiative correction.
\par
   The coupling constant $\alpha(s)$ and anomalous magnetic moment 
$a_\mu = (g-2)/2 $ of the $\mu$ are the fundamental parameter used to  
test 
the Standard Model. Fig.~\ref{fig2} shows the various sources contributing 
to the uncertainties of $a_\mu$ and $\alpha(M_z^2)$ ~\cite{blon}. It is 
obvious that the BEPC energy region plays an important role.
\par
\begin{figure}[hbt]
\begin{center}
\epsfxsize=28pc 
\fbox{See Slide9.gif} 
\caption{Relative Contributions to the Uncertainties of $a_\mu$ and 
        $\alpha(M_z^2)$     \label{fig2}}
\end{center}
\end{figure}
\par
   In order to reduce these uncertainties, BES scanned 6 points covering 
the Center-of-Mass energy from 2.6 to 5.0 GeV in 1998, and the R values 
have been published in PRL~\cite{bes2}. BES also scanned 85 points for the R 
measurement and 24 separated beam background points  in the $2.0\sim 4.8$ 
GeV energy region in 1999. All these R scan values are shown in 
Fig.~\ref{fig3}, the $\Delta R/R$ in the 2 - 5 GeV region has 
decreased from $15\% - 20\%$ (World Average) to $7\% - 10\%$.
\par
\begin{figure}[hbt]
\begin{center}
\epsfxsize=29pc 
\fbox{See Slide12.gif}
\caption{R Values Below 10 GeV  \label{fig3}}
\end{center}
\end{figure}
\par
\subsection{$J/\psi$ physics}
    $J/\psi$ decay is a good laboratory to search for 
glueballs and hybrids,  and from Fig.~\ref{fig4} we can clearly see 
that BES has the largest $J/\psi$ samples in the world, so there is an 
opportunity for BES to clarity existing problems and puzzles  
with the quark model. With the largest $J/\psi$ samples
we also can search for the Lepton Flavor Violation(LFV) and rare decays.
The analysis of $J/\psi$ decay to baryon-antibaryon final states allows
 us to study the  $N^*$ baryon, especially in the mass range of 
$1\sim 2$GeV~\cite{zou}.
\par    
\begin{figure}[hbt]
\begin{center}
\epsfxsize=27pc 
\fbox{See s44.gif}
\caption{The Distributions of the $J/\psi$ and $\psi (2s)$ Samples in the 
       world $(\times 10^6)$  \label{fig4}}
\end{center}
\end{figure}
\par
    Based on the 7.8M $J/\psi$ samples collected by BESI, we analyzed 
$J/\psi\to \gamma \gamma V (V=\rho,\phi)$ channels and found the evidence 
of the $\eta(1430)$ in both decay channels~\cite{xu}. Considering the 
branching ratios of $\eta(1430)$ in both channels, can we say whether  
it contains a glueball component?  Further investigation  
using the 50M $J/\psi$ samples collected by BESII may help us to 
understand it. 
\par
\subsection{$\psi(2S)$ physics} 
  $\psi(2S)$ is a very important field to study charmonium family 
members, and non-relativistic Perturbative QCD (PQCD).
According to PQCD: 
\begin{equation}
 \frac{B[\psi(2S) \to X_h]}{B[J/\psi \to X_h]} \approx \frac{B[\psi(2S) 
\to e^+ e^-]}{B[J/\psi \to e^+ e^-]} \approx 15\%
\end{equation}
This PQCD prediction is called the "15$\%$ Rule". But in 1984, MARKIII 
found that two channels did not obey this rule, this phenomenon called 
the "$\rho \pi$ Puzzle". 
\par
  Since BES has the largest $\psi(2S)$ sample in the 
world(Fig.~\ref{fig4}), BES has measured many channels to test the "15$\%$ 
Rule", many of them are first measurement, Fig.~\ref{fig5} 
shows the measured results which includes BES published results and 
some new preliminary results. Many of them already (or will) fill up 
the PDG and improve(or will improve) the precision.
 \par    
\begin{figure}[hbt]
\begin{center}
\epsfxsize=27pc 
\fbox{See s28.gif}
\caption{Test of the  $15\%$ Rule  \label{fig5}}
\end{center}
\end{figure}
\par
\section{The future plans for the BES and BEPC} 
   The future of BEPC and BES will be the BEPCII and BESIII. The 
BEPCII is designed to be a double ring structure, its beam energy region 
will be optimized for $1.55 \sim 1.84$ GeV, and the luminosity of BEPCII 
will achieve  about $10^{33}$ cm$^{-2}$s$^{-1}$. 
   This October, an international workshop on the BESIII detector will be 
held in Beijing, its main goals are to discuss about the design of 
BESIII.
\par
 \section*{Acknowledgments}
  We would like to thank BEPC staffs and IHEP computing center for their 
great contribution to the BES data taking. This work is supported 
by the National Natural Science Foundation of China; the Chinese Academy 
of Sciences; and the Department of Energy of USA. 

\end{document}